\documentclass[twoside]{article}
\usepackage{qic,epsfig}
\usepackage{graphicx}
\usepackage{color}
\usepackage{booktabs}
\usepackage{amsmath}
\usepackage{mathptmx}      

\begin{document}
\setlength{\textheight}{8.0truein}    

\runninghead{Maximum density of quantum information in a scalable CMOS implementation of the hybrid qubit architecture}
            {D. Rotta, M. De Michielis, E. Ferraro, M. Fanciulli and E. Prati}

\normalsize\textlineskip
\thispagestyle{empty}
\setcounter{page}{1}



\alphfootnote

\fpage{1}

\centerline{\bf
MAXIMUM DENSITY OF QUANTUM INFORMATION IN A SCALABLE}
\vspace*{0.035truein}
\centerline{\bf CMOS IMPLEMENTATION OF THE HYBRID QUBIT ARCHITECTURE}
\vspace*{0.37truein}
\centerline{\footnotesize
DAVIDE ROTTA\footnote{E-mail: davide.rotta@mdm.imm.cnr.it}}
\vspace*{0.015truein}
\centerline{\footnotesize\it Dipartimento di Scienza dei Materiali, University of Milano Bicocca}
\baselineskip=10pt
\centerline{\footnotesize\it Via R. Cozzi 53, 20126 Milano, Italy}
\vspace*{0.015truein}
\centerline{\footnotesize\it Laboratorio MDM - IMM - CNR}
\baselineskip=10pt
\centerline{\footnotesize\it Via Olivetti 2, 20864 Agrate Brianza, Italy}
\vspace*{10pt}

\centerline{\footnotesize 
MARCO DE MICHIELIS}
\vspace*{0.015truein}
\centerline{\footnotesize\it Laboratorio MDM - IMM - CNR}
\baselineskip=10pt
\centerline{\footnotesize\it Via Olivetti 2, 20864 Agrate Brianza, Italy}
\vspace*{10pt}

\centerline{\footnotesize 
ELENA FERRARO}
\vspace*{0.015truein}
\centerline{\footnotesize\it Laboratorio MDM - IMM - CNR}
\baselineskip=10pt
\centerline{\footnotesize\it Via Olivetti 2, 20864 Agrate Brianza, Italy}
\vspace*{10pt}

\centerline{\footnotesize 
MARCO FANCIULLI}
\vspace*{0.015truein}
\centerline{\footnotesize\it Dipartimento di Scienza dei Materiali, University of Milano Bicocca}
\baselineskip=10pt
\centerline{\footnotesize\it Via R. Cozzi 53, 20126 Milano, Italy}
\vspace*{0.015truein}
\centerline{\footnotesize\it Laboratorio MDM - IMM - CNR}
\baselineskip=10pt
\centerline{\footnotesize\it Via Olivetti 2, 20864 Agrate Brianza, Italy}
\vspace*{10pt}

\centerline{\footnotesize 
ENRICO PRATI
\footnote{Present address: Istituto di Fotonica e Nanotecnologia, CNR, Piazza Leonardo da Vinci 32, 20133 Milano, Italy}
}
\vspace*{0.015truein}
\centerline{\footnotesize\it Laboratorio MDM - IMM - CNR}
\baselineskip=10pt
\centerline{\footnotesize\it Via Olivetti 2, 20864 Agrate Brianza, Italy}
\vspace*{0.225truein}
\publisher{(received date)}{(revised date)}

\vspace*{0.21truein}

\abstracts{
Scalability from single qubit operations to multi-qubit circuits for quantum information processing requires architecture-specific implementations.
Semiconductor hybrid qubit architecture is a suitable candidate to realize large scale quantum information processing, as it combines a universal set of logic gates with fast and all-electrical manipulation of qubits.
We propose an implementation of hybrid qubits, based on Si Metal-Oxide-Semiconductor (MOS) quantum dots, compatible with the CMOS industrial technologic standards.
We discuss the realization of multi-qubit circuits capable of fault-tolerant computation and quantum error correction, by evaluating the time and space resources needed for their implementation.
As a result, the maximum density of quantum information is extracted from a circuit including 8 logical qubits encoded by the $[[7,1,3]]$ Steane code.
The corresponding surface density of logical qubits is 2.6 Mqubit/cm$^2$.
}{}{}

\vspace*{10pt}

\vspace*{3pt}
\communicate{to be filled by the Editorial}

\vspace*{10pt}\textlineskip    

\section{Introduction}
\label{intro}
\noindent

Several implementations have been considered for circuital quantum information processing (QIP) \cite{QIP_DiVincenzo,Review_Ladd,Review_BulutaNori,Review_SiQIP_Morton}.
Nevertheless, scalability is still an issue for many architectures. 
In this framework, a Complementary Metal-Oxide-Semiconductor (CMOS) architecture based on silicon would take full advantage of the well-known physical properties of the material and of the mature technological improvements driven by the semiconductor industry. 
Here we provide a CMOS-compatible architecture for QIP in silicon and evaluate the performances of the fundamental logic gates and the physical constraints for their scalability to multi-qubit logic circuits.
As a result, two important parameters are determined for this implementation: the maximum surface density of quantum information and the characteristic time for quantum communication between two logic qubits.

Implementations of charge and spin qubits in silicon have been explored in both quantum dots sytems \cite{Review_SiQIP_Morton,MDM_APEX,Prati_Nanotech,Morello_Nature2013_QD}
and single dopant atom transistors \cite{Kane,Yablonovich_SiGe_Qubit,Hollenberg_Charge,KoillerSaraiva_SiQIP,Leti_APL,Mazzeo_APL,Varenna,Prati_Nature,Morello_NanoLett}. Coherent manipulation of quantum states have been demonstrated in both atomic systems \cite{Morello_T1,Morello_T2}, as well as in semiconductor quantum dots (QD), which take advantages from more relaxed bounds on the device dimensions \cite{LossDiVinc,Spin_Qubits_Kloeffel}.
Besides single spin and charge qubits \cite{Koppens,LZS_Charge_Qubit}, that make use of a single electron in a double quantum dot (DQD), in the last decade several architectures have been explored by employing two electron spins (S-T$_0$ $i.e.$ singlet-triplet qubit) \cite{Petta,Yacoby_2qubit} or three spins in double \cite{Shi_Hybrid} and triple QDs \cite{DiVinc_Marcus_3QD}. 
Although satisfactory results have been achieved mainly in III-V heterostructures, spin-orbit coupling and hyperfine interactions are weaker in silicon \cite{Tyryshkin_T2_Seconds,Review_SiQIP_Morton,Morello_Nature2013_QD}, suggesting that silicon itself could be a promising platform for QIP. 
Focusing on silicon DQD qubits, the controlled manipulation of qubit states has been obtained for a S-T$_0$ qubit \cite{HRL} and a hybrid qubit \cite{Shi_Nature_Hybrid,Hybrid_2014} in SiGe heterostructures. 
In particular, the hybrid qubit is an attractive candidate for a large scale integration of QIP, as it allows a fast and all-electrical manipulation of the qubit states, with no need for either a strong magnetic field gradient, like in S-T$_0$ qubits, or microwave antennas, required in single spin qubits. 
Besides the study of the electronic properties of Si-MOS QDs for QIP \cite{Pierre_APL,Prati_Nanotech,MDM_APEX}, we also derived the effective hamiltonian for hybrid qubits, defining the pulse sequences to perform universal quantum computation with such architecture \cite{Ferraro_QIP,LavoroLungo,Universal_Set}. 
Here we calculate the maximum storage of quantum information processing allowed by a CMOS-compatible implementation of silicon hybrid qubits.

In Section \ref{sec:Hybrid qubits} the logic basis of the hybrid qubit is presented, as well as the operation of 1-qubit and 2-qubit logic gates for universal QIP.
In Section \ref{Technology} their feasibility is discussed in a state of the art CMOS process: the physical requirements for the manipulation of quantum states are compared with the constraints imposed by the existing technologies and realistic devices are designed to implement data and communication qubits.
Finally, in Section \ref{LSI} the large scale integration of silicon hybrid qubits is considered in multi-qubit networks capable of fault tolerant computation and quantum error correction.
The maximum surface density of logic qubits per unit area is estimated, as well as the time load for quantum communication between two logic qubits.



\section{Fundamental logic gates in the hybrid qubit architecture}
\label{sec:Hybrid qubits}
\noindent

This Section is devoted to the description of the main concepts underliying the hybrid qubit architecture.
The main building blocks for such architecture are defined in terms of data and communication qubits.
Data qubits perform quantum information processing, that is one and two qubit logic operations as well as initialization and read-out of individual qubit states.
Communication qubits, conversely, are devoted to the communication of quantum information between distant data qubits.

In Subsection \ref{subsec:data_qubits} the fundamentals of hybrid architecture are introduced, as well as the schematic design and the operation of the quantum logic gates with one and two interacting qubits. In Subsection \ref{subsec:comm_qubits} we show how quantum information can be transmitted between distant hybrid qubits through the sequential repetition of SWAP logic gates between adjacent communication qubits.

\subsection{Data qubits: one and two qubit logic gates}
\label{subsec:data_qubits}
\noindent
An architecture that promises the best compromise among fabrication, fast gate operations, manipulation and scalability is the hybrid qubit proposed in Refs. \cite{Shi_Hybrid,Shi_Nature_Hybrid,Hybrid_2014,Ferraro_QIP,LavoroLungo,Universal_Set,Koh_PNAS}. It consists of two quantum states based on three electrons electrostatically confined in two QDs, with at least one electron in each. The convenience to use such an architecture is due to the possibility of obtaining fast gate operations with purely electrical manipulations. The exchange interaction, which is the dominant mechanism of interaction between adjacent spins, suffices for all the one and two qubits operations. In addition the three electrons spin system removes the need of using oscillating magnetic or electric fields or quasi-static Zeeman field gradient to realize full qubit control, which is required for instance in singlet-triplet qubits \cite{Petta}. Starting from an Hubbard-like model we have derived in Ref. \cite{Ferraro_QIP} a general effective Hamiltonian for the hybrid qubit in terms of only exchange interactions among the three electrons. 

To define the logic basis for the hybrid qubit let's first introduce some preliminary notions. The total Hilbert space of three electron spins has a dimension of 8 and the total spin eigenstates form a quadruplet with $S=3/2$ and $S_z=-3/2;-1/2;+1/2;+3/2$ and two doublets each with $S=1/2$ and $S_z=\pm1/2$, where the square of the total spin is $\hbar^2S(S + 1)$ and the z-component of the total spin is $\hbar S_z$. The qubit is encoded in the restricted two-dimensional subspace with spin quantum numbers $S=1/2$ and $S_z=-1/2$, like in Ref. \cite{Shi_Hybrid}. We point out that only states with the same $S$ and $S_z$ can be coupled by spin independent terms in the Hamiltonian. The logic basis $\{|0\rangle,|1\rangle\}$ used is constituted by singlet and triplet states of a pair of electrons in combination with the angular momentum of the third spin, that is:
\begin{equation}\label{01}
|0\rangle\equiv|S\rangle|\downarrow\rangle, \qquad |1\rangle\equiv\sqrt{\frac{1}{3}}|T_0\rangle|\downarrow\rangle-\sqrt{\frac{2}{3}}|T_-\rangle|\uparrow\rangle
\end{equation}
where $|S\rangle$, $|T_0\rangle$ and $|T_-\rangle$ are respectively the singlet and triplet states
\begin{equation}
|S\rangle=\frac{|\uparrow\downarrow\rangle-|\downarrow\uparrow\rangle}{\sqrt{2}}, \quad |T_0\rangle=\frac{|\uparrow\downarrow\rangle+|\downarrow\uparrow\rangle}{\sqrt{2}}, \quad |T_-\rangle=|\downarrow\downarrow\rangle
\end{equation}
in the left dot, and $|\uparrow\rangle$ and $|\downarrow\rangle$ respectively denote a spin-up and spin-down electron in the right dot.

Every logic operation starts with the initialization process, when all the variables are regulated through appropriate external electric and magnetic fields \cite{Shi_Hybrid}. During this procedure, all the qubits composing the system are moved in the state corresponding to the 0 logic state. Starting from this condition it is possible to proceed further with the operations that are generally described by unitary matrices that finally lead to the desired logic gates. 

\begin{figure}[ht]
\centerline{\includegraphics[width=1.0\textwidth]{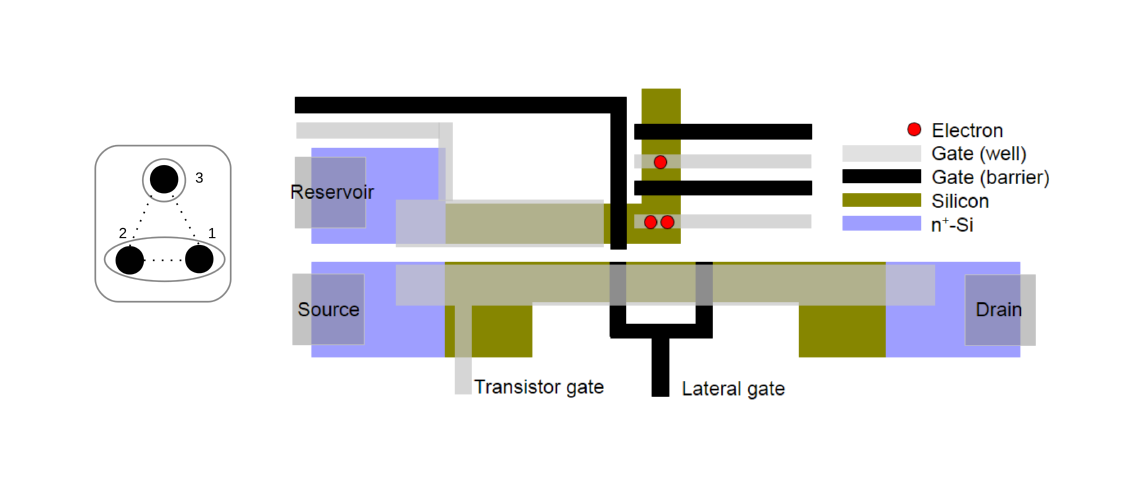}}
\vspace*{13pt} \fcaption{Left: schematic of the configuration for the hybrid qubit; electrons are denoted by 1, 2 and 3; dotted lines indicate the main interactions. Right: qualitative design of the device holding a single hybrid qubit with the related \emph{reservoir} and SET. The three electrons, highlighted as red circles, are electrostatically confined in the double QD by means of metallic gates (well and barrier). The electron \emph{reservoir} is added to allow the read-out of the qubit state through the SET.}\label{Fig:device} 
\end{figure}
In the following, an implementation of the hybrid qubit is presented. A sketch of the device is reported in Figure \ref{Fig:device}, where metal gates form two electrostatic QDs and control the energy barrier between them. However, additional structures are needed to inject electrons in the QDs. This can be achieved by fabricating a \emph{reservoir} as source of electrons near the double QD and by controlling the height of an energy barrier between the \emph{reservoir} itself and the double QD through an electrostatic gate. In addition, the fabrication of a charge sensor is needed for the readout of the qubit state which coincides with the read out of the spin state of electrons confined in the doubly occupied QD. To serve this purpose, a Single Electron Transistor (SET), which is a MOSFET where a QD is formed by placing additional lateral gates orthogonal to the channel, can be used to electrostatically sense the spin state of the electrons in the doubly occupied QD.  

Once that the operations on the qubits are concluded, the next step is represented by the read out process, as is described in the following. When read out of the qubit starts, tunneling is allowed from the doubly occupied QD to the \emph{reservoir} by a reduction of the interposed electrostatic barrier. When the electron pair is in a singlet state the corresponding wavefunction is more confined and the tunneling rate to the \emph{reservoir} is lower than that of the triplet state, which has a broader wavefunction. When the electron tunnels, the electrostatic potential landscape changes and so does the current passing through the electrostatically coupled SET. The measurement of the time interval between the read out signal and the current variation in the SET is supposed to reveal the spin state of the electron pair \cite{Shi_Hybrid}.

\begin{figure}[ht]
\centerline{\includegraphics[width=1.0\textwidth]{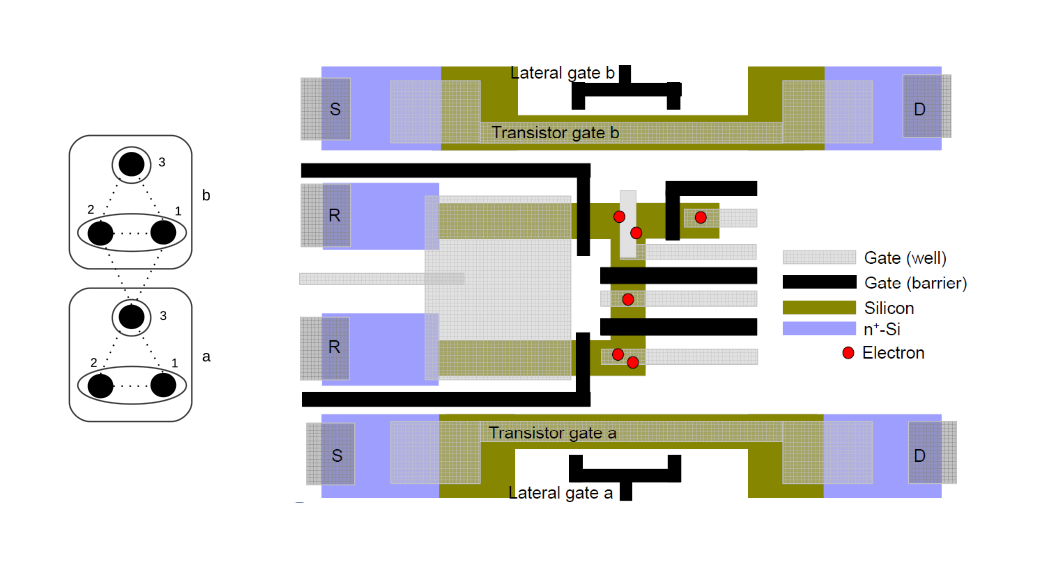}}
\vspace*{13pt} \fcaption{Left: schematic of the configuration for the couple of interacting hybrid spin qubits denoted by a and b; electrons are denoted by 1, 2 and 3; dotted lines indicate the main interactions. Right: qualitative design for a couple of interacting hybrid qubits.}
\label{Fig:deviceCNOTs} 
\end{figure}
By adopting the same approach as for the single-qubit logic gates, the extended effective Hamiltonian model for two interacting qubits is derived \cite{LavoroLungo}. The total system, composed by six electron spins, is described within the subspace with total angular momentum operator $S=1$ and $S_z=-1$ adopting the basis $\{|00\rangle, |01\rangle, |10\rangle,|11\rangle\}$, where the logic state $|0\rangle$ and $|1\rangle$ are defined in Eq.(\ref{01}). 
A possible layout for two hybrid qubit gates is sketched in Figure \ref{Fig:deviceCNOTs}, where two data qubits are put in close connection by a controllable electrostatic barrier.
The Controlled-NOT (CNOT) gate, for example, is obtained by using the sequence reported in Refs. \cite{LavoroLungo,Universal_Set}.
The gates in Figure \ref{Fig:device} and \ref{Fig:deviceCNOTs} are sufficient to carry out arbitrary qubit rotations as well as general two qubit operations, providing a complete set of quantum gates for universal quantum computing.

\subsection{Communication qubits: the SWAP chain}
\label{subsec:comm_qubits}
\noindent

In this paragraph the problem of the communication among data qubits within an interconnecting circuit is analysed and an efficient strategy for an optimal transfer is presented. The model is based on hybrid qubits chains where the exchange interaction is exploited to transfer end to end the logic states through SWAP operations, where the SWAP gate operates an exchange between the states of two adjacent qubits. 

\begin{figure}[t!]
\begin{center}
\includegraphics[width=0.6\textwidth]{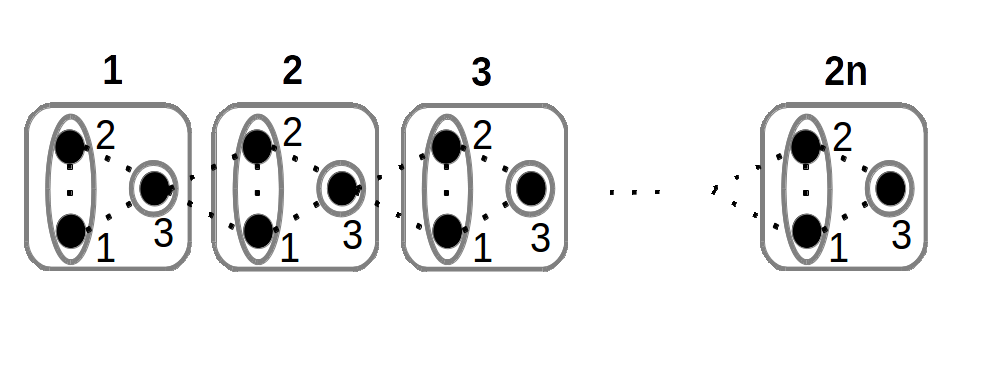}
\includegraphics[width=0.6\textwidth]{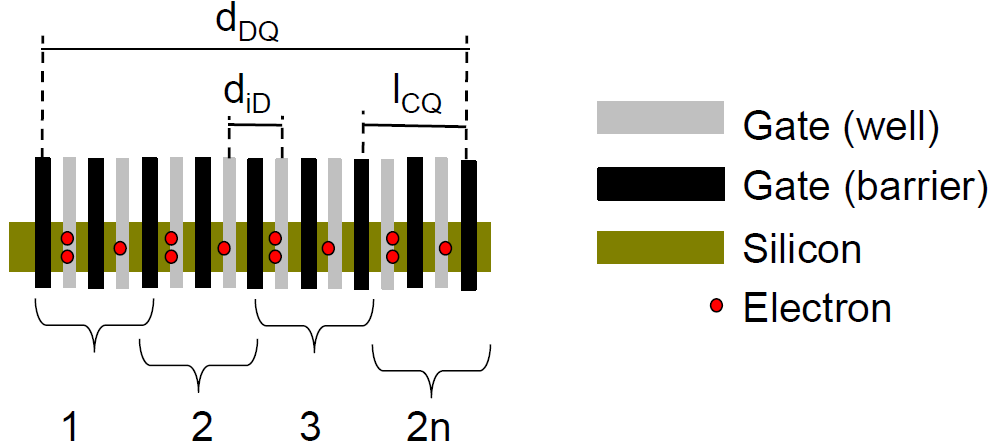}
\end{center}
\vspace*{13pt} \fcaption{Top: schematic of a chain of an even number ($2n$) of hybrid qubits. Bottom: example of qualitative design for the implementation of quantum dot chains. The electrons confined in the QDs are highlighted in red. 
$d_\text{iD}$ is the inter-dots distance, $l_\text{CQ}$ = $2d_\text{iD}$ is the length of the communication qubit and $d_\text{DQ}$ is the distance between the interconnected data qubits which corresponds to the distance between the head and the tail of the chain.}\label{Fig:deviceSWAPChain} 
\end{figure}

In Figure \ref{Fig:deviceSWAPChain} a scheme and a qualitative design of the hybrid qubit chain is shown where an even number of hybrid qubits are put into direct connection. 

\begin{figure}[h]
\centerline{\includegraphics[width=0.7\textwidth]{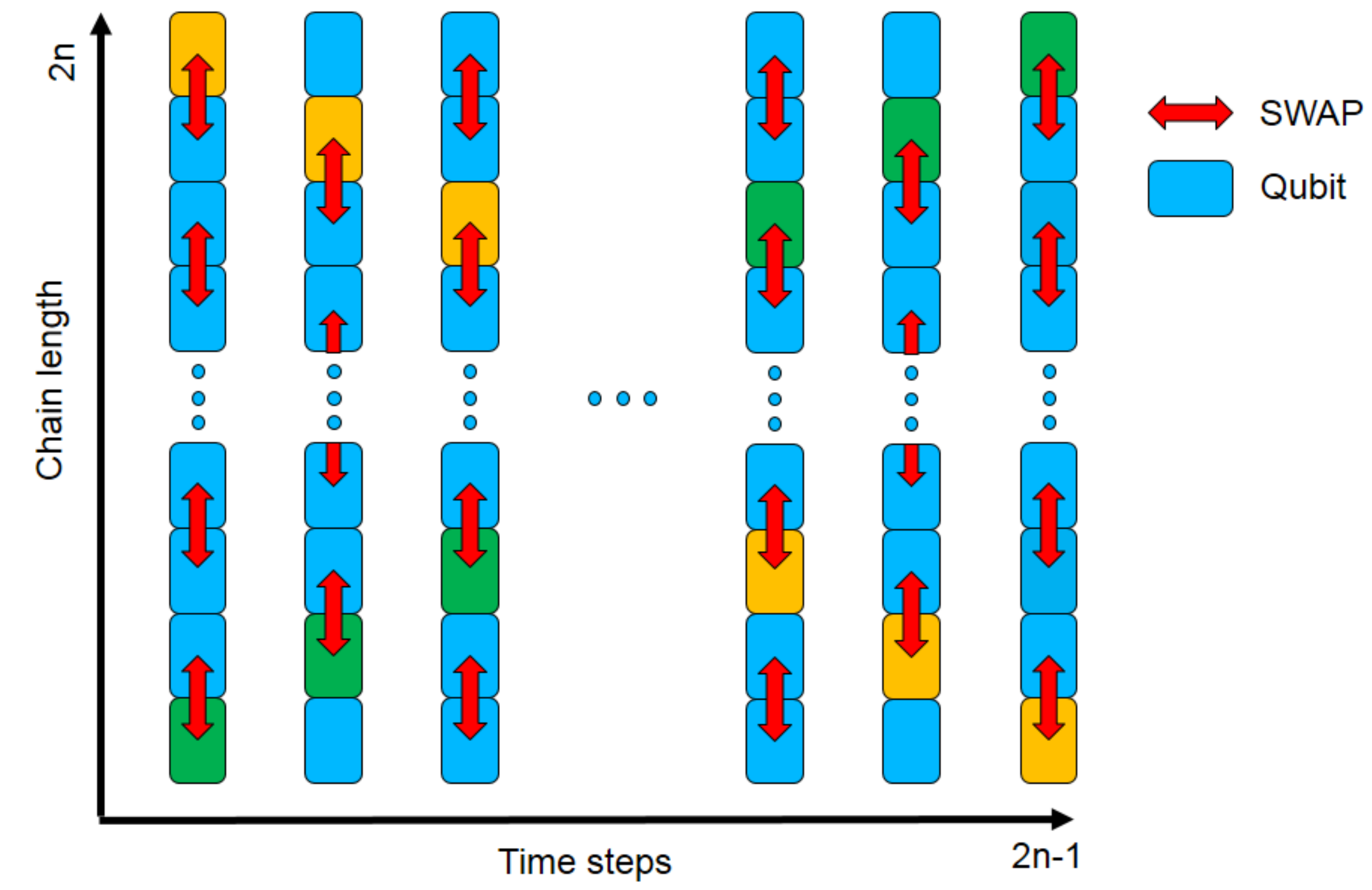}}
\vspace*{13pt} \fcaption{State representation of the hybrid qubit chain as a function of time. A sequence of $2n-1$ SWAP steps, with time duration $t_{SWAP}$, is applied when a chain of $2n$ hybrid qubits is considered. SWAP operations between qubits 1-2, 3-4, 5-6, etc. are required in odd time steps whereas qubit 2-3, 4-5, 6-7, etc. are swapped in even time steps. 
As a result, an independent control of all the gates in each SWAP step is not required. Gates can be grouped in two sets and driven alternatively, making the chain control easier. After $2n-1$ SWAP steps, a bidirectional transfer of the states initially localized at the extremities of the chain is obtained.
}\label{Fig:SWAPchain} 
\end{figure}

From a practical point of view it is necessary firstly to initialize each qubit, which is composed by $3\times2n$ electrons (3 is the number of electrons for each qubit and $2n$ è is the total number of hybrid qubits).
The transfer begins when the state of the system at the head of the chain has been exchanged through a SWAP operation with the state of the adjacent qubit. In this way the first qubit has received the state of the second qubit and vice versa. At the second step the same mechanism involves the second and the third qubit. At the end of the process, the information has been completely transferred to the last qubit. 
In this case the exchange is operated sequentially. An optimized control, as depicted in Figure \ref{Fig:SWAPchain}, will allow to operate the exchange in parallel.
It is possible to operate the exchange in parallel instead of in sequence, optimizing the bidirectional transfer of the states as pictured in Figure \ref{Fig:SWAPchain}.

\begin{figure}[h!]
\centerline{\includegraphics[width=0.8\textwidth]{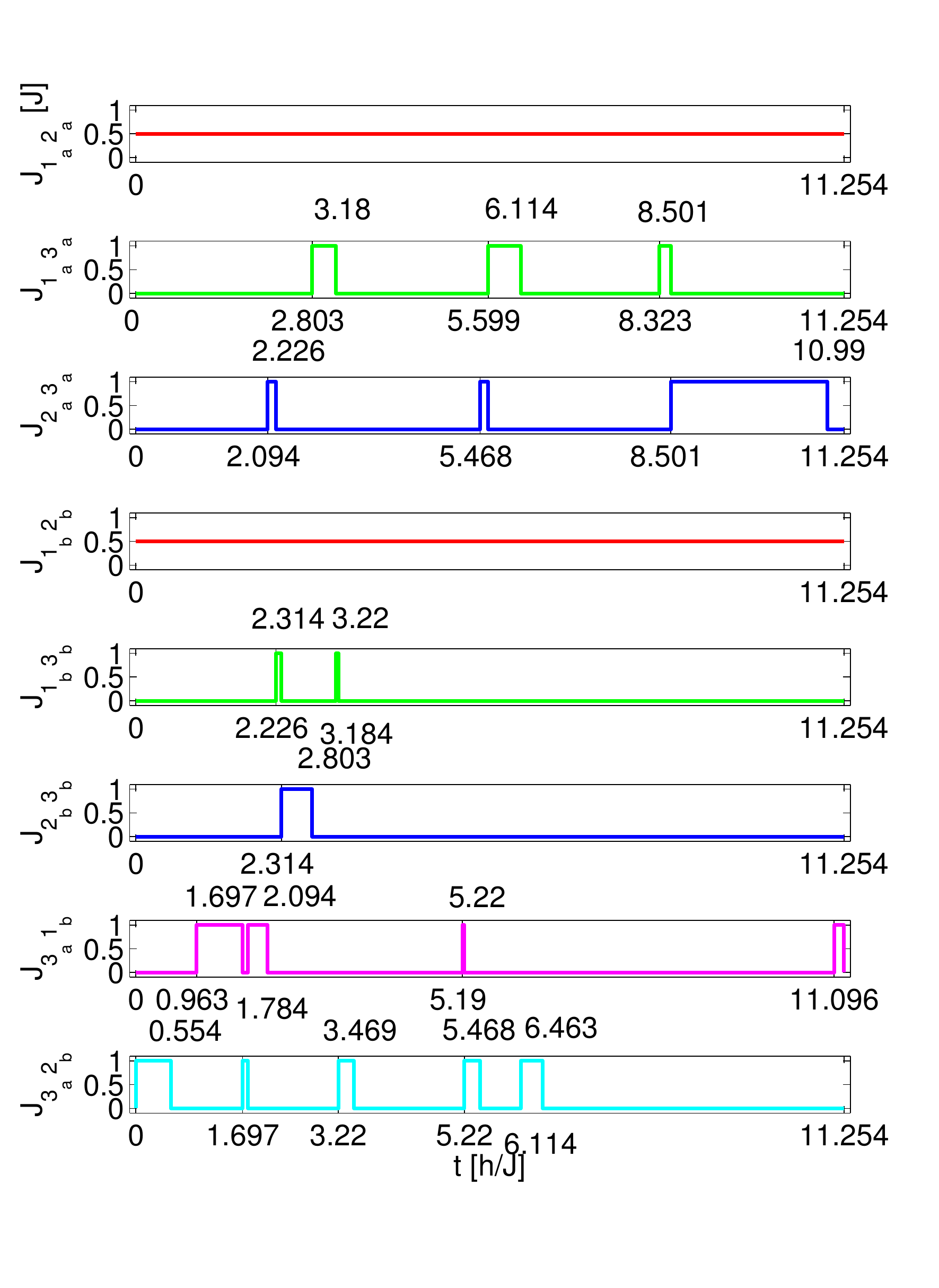}}
\vspace*{13pt} \fcaption{Waveforms of the effective exchange variables implementing a SWAP gate (up to a global phase) with fixed $J_{12}$ = $J$/2 in both qubits. Times are in unit of $h/J$.}\label{Fig:SeqSWAPv1} 
\end{figure}

In order to find the gate sequences necessary to generate a SWAP operation between two qubits we employed the same search algorithm used in \cite{LavoroLungo}. The resulting pulse sequence to obtain the SWAP gate is reported in Figure \ref{Fig:SeqSWAPv1}, where the different $J$ terms represents the exchange parameters \cite{LavoroLungo}.

Only a couple of electrons interact after tuning the tunneling parameters between the dots belonging to the same qubit or to different ones. 
Interactions between $1_{a}$ and $2_{a}$ and between $1_{b}$ and $2_{b}$ have been set to a constant value in the search algorithm, as they are not effectively manipulable from the external \cite{LavoroLungo}.
More in detail, $J_{1a2a}$ = $J_{1b2b}$ = $J$/2 where J is the maximum effective exchange interaction between the two dots.

\begin{figure}[ht]
\centerline{\includegraphics[width=0.75\textwidth]{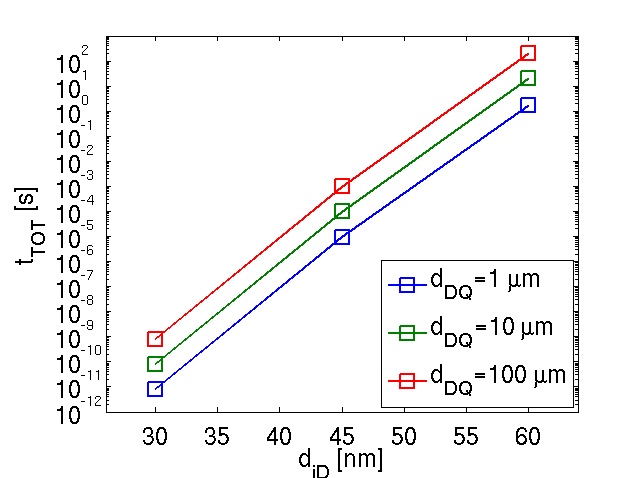}}
\vspace*{13pt} \fcaption{Graph of the total time SWAP chain $t_\text{TOT}$ as a function of the inter QD distance $d_\text{iD}$ for three different distances $d_\text{DQ}$ between head and tail data qubits.}\label{Fig:t-dint} 
\end{figure}

In order to quantitatively design the qubit chain, the simulation results on the single qubit reported in Ref. \cite{LavoroLungo} are used.
The SWAP time, $t_\text{SWAP}$, between the states of two adjacent hybrid qubits depends on the exchange interaction $J$ that again depends on the tunneling rate $t_r$ between the energy levels in the two QDs forming the qubit.
$t_r$ depends on the inter-dots distance $d_\text{iD}$ and it is linked to the length of the communication qubit by $l_\text{CQ}$ =$2d_\text{iD}$. The number of qubits $2n$ forming the qubit chain depends on the ratio between the head to tail distance between data qubits $d_\text{DQ}$ and $d_\text{iD}$. The total time $t_\text{TOT}$ to transfer the information from one extremity to the other by successive SWAP operations is:
\begin{equation}
t_\text{TOT} = (2n-1) \cdot t_\text{SWAP}= \Big(\frac{d_\text{DQ}}{l_\text{CQ}} -1 \Big) \cdot t_\text{SWAP}(d_\text{iD}) = \Big( \frac{d_\text{DQ}}{2d_\text{iD}} -1 \Big) \cdot t_\text{SWAP}(d_\text{iD})
\end{equation}
where $t_\text{SWAP}=t_\text{seq} \cdot h/J$ and $t_\text{seq}$ is the duration of the sequence in units of $h/J$.
$J$ is estimated with $J=t_r^{2}/\Delta E_\text{ST}$ where $\Delta E_\text{ST}$ is the singlet-triplet energy splitting \cite{Shi_Hybrid}.
In Figure \ref{Fig:t-dint} the total chain time $t_\text{TOT}$ is reported as a function of $d_\text{iD}$ for three different values of $d_\text{DQ}$.
Total time $t_\text{TOT}$ increases exponentially as $d_\text{iD}$ raises.

\section{CMOS implementation of the hybrid qubit architecture}
\label{Technology}
\noindent

In this paragraph we explore the limiting size of the discrete components to implement the hybrid qubit architecture imposed by a CMOS-compatible fabrication process.

The standard in semiconductor industry is set by silicon CMOS manufacturing, due to the capability to fabricate p- and n-channel devices on the same chip and to build devices with a low power consumption \cite{Nishi}. 
Hence, the technologic constraints set by the 22 nm technologic node of CMOS nanoelectronics are examined in Section \ref{subsec:process}.
According to such physical constraints, realistic data and communication qubits are designed and reported in Section \ref{subsec:masks}, providing the building blocks for a complete implementation of the hybrid qubit architecture in silicon.

\subsection{Technologic constraints of CMOS manufacturing at the 22 nm node}
\label{subsec:process}
\begin{figure}[h!]
\centering
  \includegraphics[width=0.8\columnwidth]{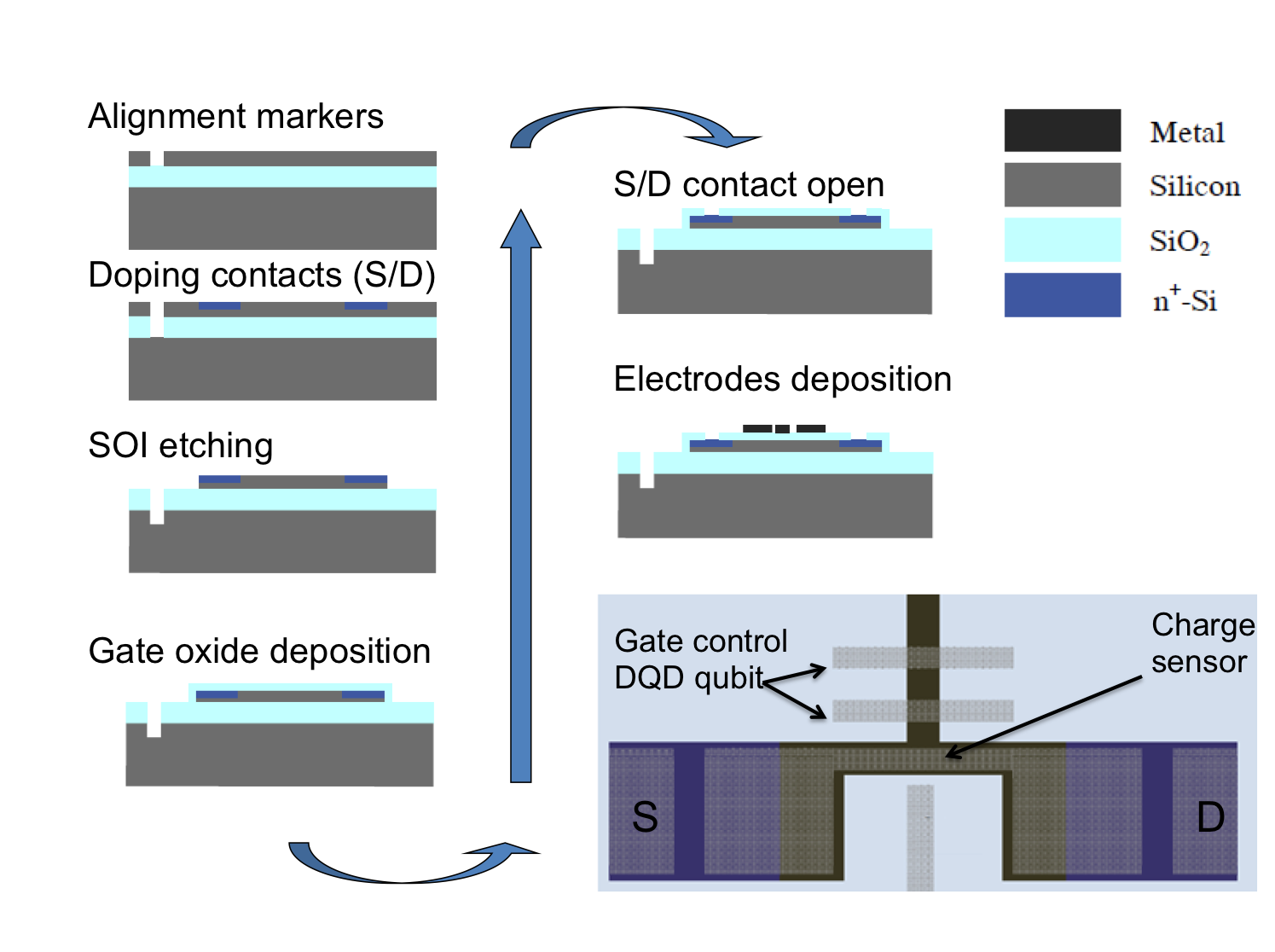}
\vspace*{13pt} \fcaption{Schematic process flow for a CMOS-compatible realization of semiconductor hybrid qubits on SOI wafers.
A schematic top view of the resulting device is shown on the right bottom.}
\label{fig:process}       
\end{figure}

Figure \ref{fig:process} shows a schematic process flow for the realization of Si-MOS hybrid qubits on a Silicon-On-Insulator (SOI) platform.

Quantum gates are defined by selective etching of the SOI device layer after a lithographic patterning exposure.
Then a gate insulator, for example Al$_2$O$_3$ or other high-$k$ oxides, is deposited on the silicon islands \cite{Nishi,ITRS}. Finally, the metal electrodes for source-drain leads and the electrostatic gates are deposited and patterned by a further lithographic exposure.
The mentioned process flow can be completely adapted to an industrial one: all the steps can be performed with the main deposition and etching techniques employed in a common industrial production line, such as Chemical Vapor Deposition (CVD), Atomic Layer Deposition (ALD) and Reactive Ion Etching (RIE) \cite{Nishi}.

The main critical issues concern the lithographic steps, because a few nanometer resolution is required. %
At the present times, most of the patterning processes in microelectronics and micromachining for Ultra Large Scale Integration (ULSI) are carried out through Deep Ultra-Violet (DUV) lithography, that makes use of ArF laser sources ($\lambda$ = 193 nm) and is capable of high resolution as well as high throughput (100 wafers per hour) \cite{Nishi,ITRS}. 
According to the last International Technology Roadmap for Semiconductors (ITRS) Update, 22 nm is the minimum half-pitch of un-contacted Poly-Si in flash memories and it represents a significative benchmark of the ultimate resolution of DUV lithography at the present node \cite{ITRS}. 
Further improvements are expected in the very next years to reach the next technological node set at 16 nm \cite{ITRS}. 
In this perspective, alternative lithographic techniques, like Extreme Ultra-Violet (EUV) lithography and multi-beam Electron Beam Lithography (EBL), are examined to push technology to the forthcoming nodes  \cite{ITRS}.

A minimum feature size of 20 nm is a reasonable design rule for a realistic implementation of the hybrid qubit architecture proposed in Section \ref{sec:Hybrid qubits}.
All the masks for data and communication qubits have been designed accordingly and are compatible wih the 22 nm node technology.

\subsection{Realistic design of data and communication qubits}
\label{subsec:masks}
\begin{figure}[ht]
\centering
\includegraphics[width=\columnwidth]{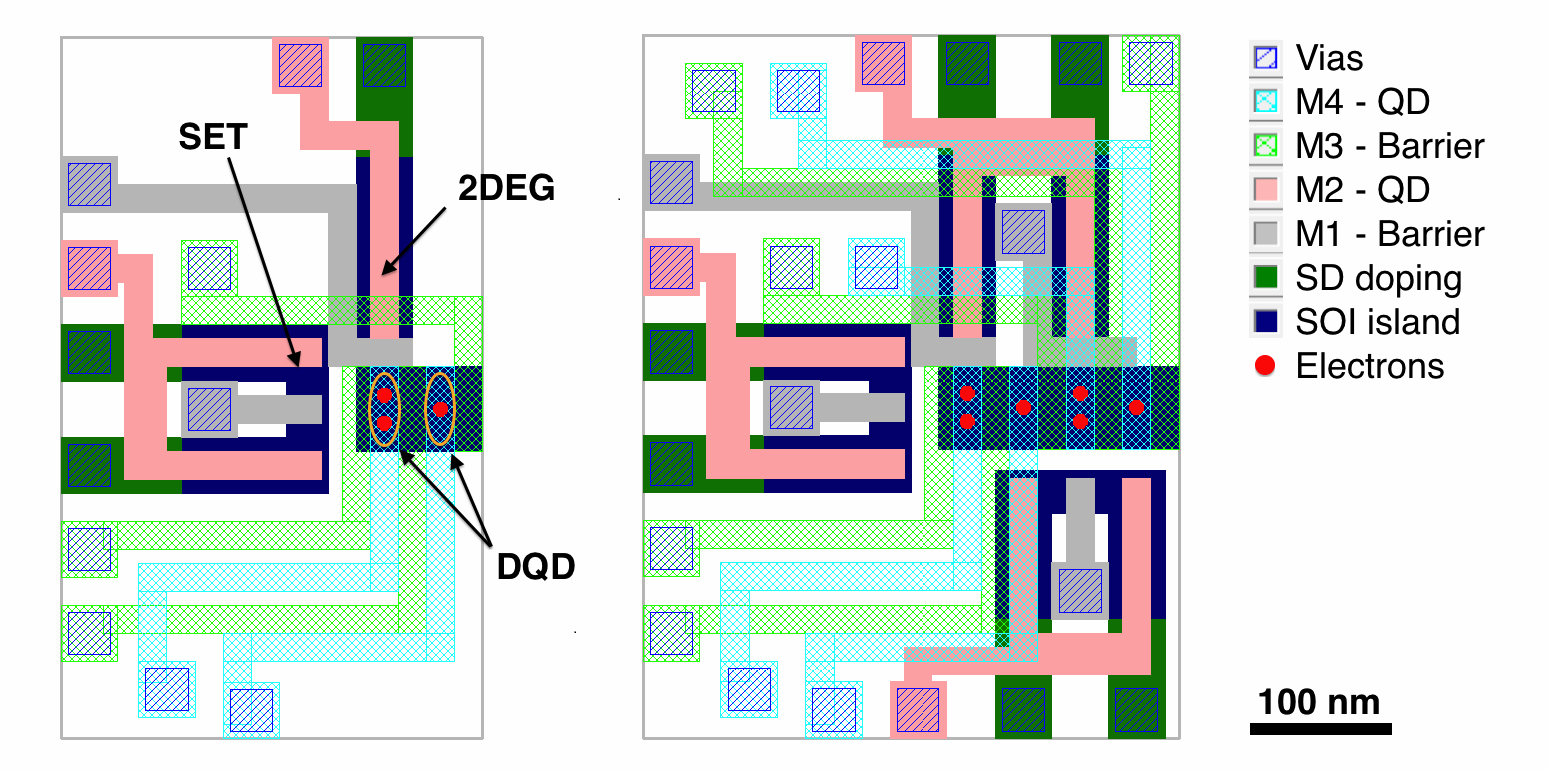}
\vspace*{13pt} \fcaption{Lithographic masks for data qubits.
Left: 1-qubit gate is composed of a DQD contacted with an electronic \emph{reservoir} (2DEG) for the system initialization and read-out and a SET for charge sensing of the DQD.
Right: 2-qubit gate. The mask is designed analogously to the 1-qubit gate to obtain full control over two independent qubits coupled by a tunable electrostatic barrier.
A color code on the right side identifies the lithographic levels required for the fabrication process, corresponding respectively to the definition of silicon islands, degenerate doping at the source and drain contacts, four levels for metal gates and vias for electrical connections.
The minimum feature size is 20 nm for both masks, whereas the total area is 300 x 500 nm$^2$ for one qubit and 380 x 500 nm$^2$ for two qubits.}
\label{fig:qubit}       
\end{figure}

\noindent
The mask for a single hybrid qubit is reported on the left of Figure \ref{fig:qubit}. 
One lithographic level, that defines the SOI islands, is in blue, while the other levels correspond to four superposed metal levels for the electrical contacts, high doping for the electron \emph{reservoirs} and vias to Back End Of Line (BEOL) levels.
In Figure \ref{fig:qubit} we highlighted the silicon regions where the DQD qubit and SET charge sensor are defined.
Grey and green electrodes (level 1 and 3) are used as inter-dot barriers, while red and light blue ones (level 2 and 4) act as plunger gates defining the potential wells and controlling the chemical potential in the quantum dots.
The minimum feature size for structures on the same level is 20 nm.
As a result, the one-qubit gate in Figure \ref{fig:qubit} can be realized on a submicrometer area (300 x 500 nm$^2$). 

A two-qubit gate can be obtained as a replica of the one-qubit mask with few adjustments.
The mask on the right of Figure \ref{fig:qubit} covers an active area of 380 x 500 nm$^2$ and it's composed of two DQDs with separated electron \emph{reservoirs} and SET charge sensors for independent initialization and read-out of the two qubits.
The doubly occupied dot is closer to the charge sensor in both qubits and can directly communicate with the electron \emph{reservoir} to facilitate the read-out procedure. 
The gates in Figure \ref{fig:qubit} are sufficient to carry out arbitrary qubit rotations as well as general two qubit operations, like the Controlled-NOT (CNOT).
As a results, such devices provide a complete set of logic gates and represent the starting point to perform universal quantum computation with the hybrid qubit architecture.

\begin{figure}[ht]
\centering
  \includegraphics[width=0.7\columnwidth]{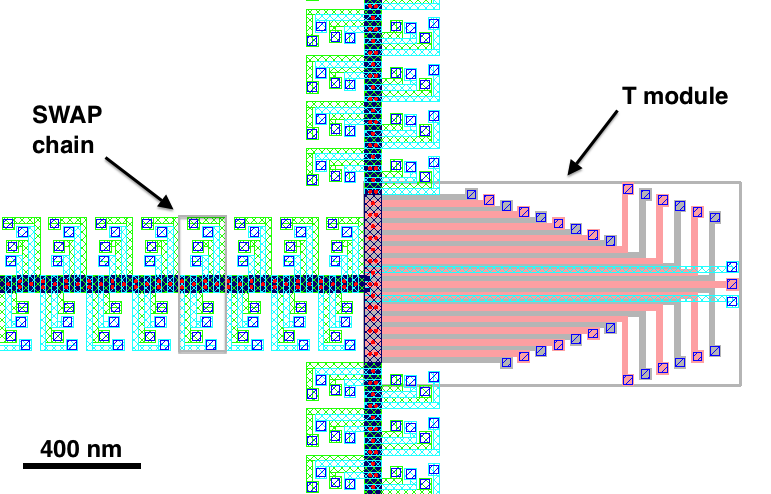}
\vspace*{13pt} \fcaption{Communication qubits for the coherent transfer of quantum information in a 2-dimensional array of qubits.
Chain modules enable quantum communication between adjacent qubits through a SWAP logic operation. The T module, on the right side, is a modified version of the chain module that makes it possible to bring in contact orthogonal qubit chains.
The lithographic levels are depicted according to the color code in Figure \ref{fig:qubit}.
The shaded squares indicate the active area of the chain module (160 x 460 nm$^2$) and of the T module (1300 x 700 nm$^2$).}
\label{fig:chain}       
\end{figure}

According to the discussion in Section \ref{sec:Hybrid qubits}, quantum communication can be accomplished by sequential SWAP operations across a qubit chain.
We designed two modular structures, namely the "chain" and the "T" module, to be composed in arbitrary 2-dimensional arrays of communication qubits (Figure \ref{fig:chain}).
The chain module consists of two qubits controlled by independent plunger and barrier gates with the only scope to carry out a SWAP operation between the states of the two qubits with no need to initialize and read-out the logic states.
Analogously, the T-module is a modified and rearranged version of a multi-chain module and acts like a crossroad for flying qubits: orthogonal qubit chains are brought in contact through this module, creating the conditions for 2-dimensional arrays of qubits.


\section{Quantum computing on a large scale}
\label{LSI}
\noindent

In this Section a possible integration on a large scale of silicon hybrid qubits is evaluated and the maximum quantum information density per unit surface is estimated.
The occurrence of faulty logic gates and memory errors is taken into account and a Quantum Error Correction scheme is proposed to improve the effective gate fidelity in multi-qubit circuits.
In this framework, two important figure of merit are estimated: the maximum density of logic qubits per unit area and the time for quantum communication between logic qubits.

Gate fidelity is an important metrics in a QIP architecture, since errors are much more frequent in quantum computers than in their classical counterparts.
In fact, information in qubits is rapidly corrupted by decoherence, $i. e.$ the interaction with the sorrounding environment.
In hybrid qubits charge and spin noise are the main sources of decoherence and they induce unwanted rotations in the Bloch sphere.
As a result, charge and spin noise are responsible for errors with a probability of about $10^{-3}$ errors per logic gate \cite{Koh_PNAS}.
Generally, protection against errors in QIP is achieved through bit encoding and fault-tolerant computation \cite{QEC_Beginners}.

\begin{figure}[t!]
\centering
  \includegraphics[width=0.9\columnwidth]{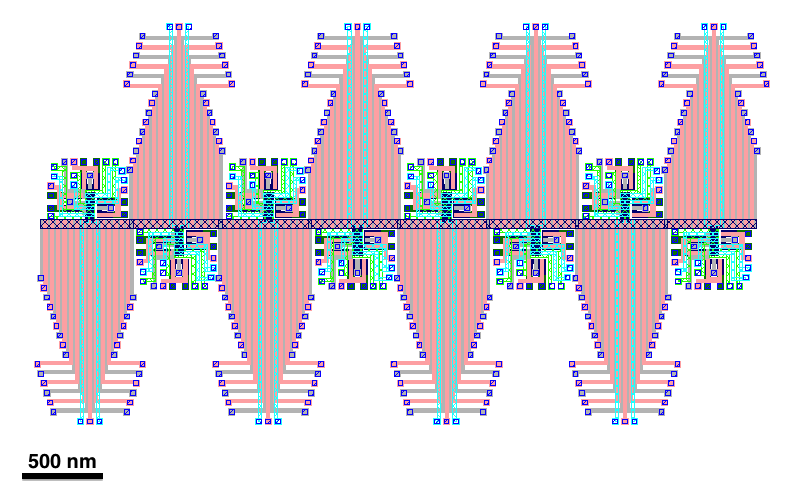}
\vspace*{13pt} \fcaption{Design of a logic qubit encoded by the $[[7,1,3]]$ Steane code in 7 physical qubits.
Lithographic levels are represented according to the color code in Figure \ref {fig:qubit}.
This multi-qubit circuit is composed of 8 data qubit gates (the 2-qubit device reported on the right of Figure \ref{fig:qubit}) and 8 T modules for quantum communication (presented in Figure \ref{fig:chain})
The minimum feature size is 20 nm, while the total area is 11.642 $\mu$m$^2$.}
\label{fig:qubyte}       
\end{figure}

A multi-qubit circuit is considered to this extent in Figure \ref{fig:qubyte}, where many data qubits are connected by T-modules on a bus structure. 
Here a logic qubit can be encoded by 7 physical qubits according to the $[[7,1,3]]$ Steane code, allowing for fault-tolerant computation and quantum error correction (QEC) \cite{QEC_Beginners,QEC_Steane}. 
In this framework, the information of a logic qubit is stored in a 2-dimensional subspace of the $2^7$-dimensional Hilbert space defined by 7 physical qubits.
As a result, a logic qubit is less susceptible to single physical qubit failures, since a faulty gate can be revealed and corrected with standard QEC techniques maintaining the coherence of the logic qubit.
Actually, QEC algorythms require some supplementary qubits for the measurement of the syndrome of the logic qubit.
The number of such extra-qubits, or \emph{ancillae}, generally depends on the quantum code, the QIP architecture and the specific procedure of error detection and correction.
In particular, 12 auxiliary qubits are sufficient for a complete QEC algorithm with the $[[7,1,3]]$ code \cite{QEC_Beginners,IEEE_SiP_Arch}.

\begin{figure}[ht]
\centering
  \includegraphics[width=\columnwidth]{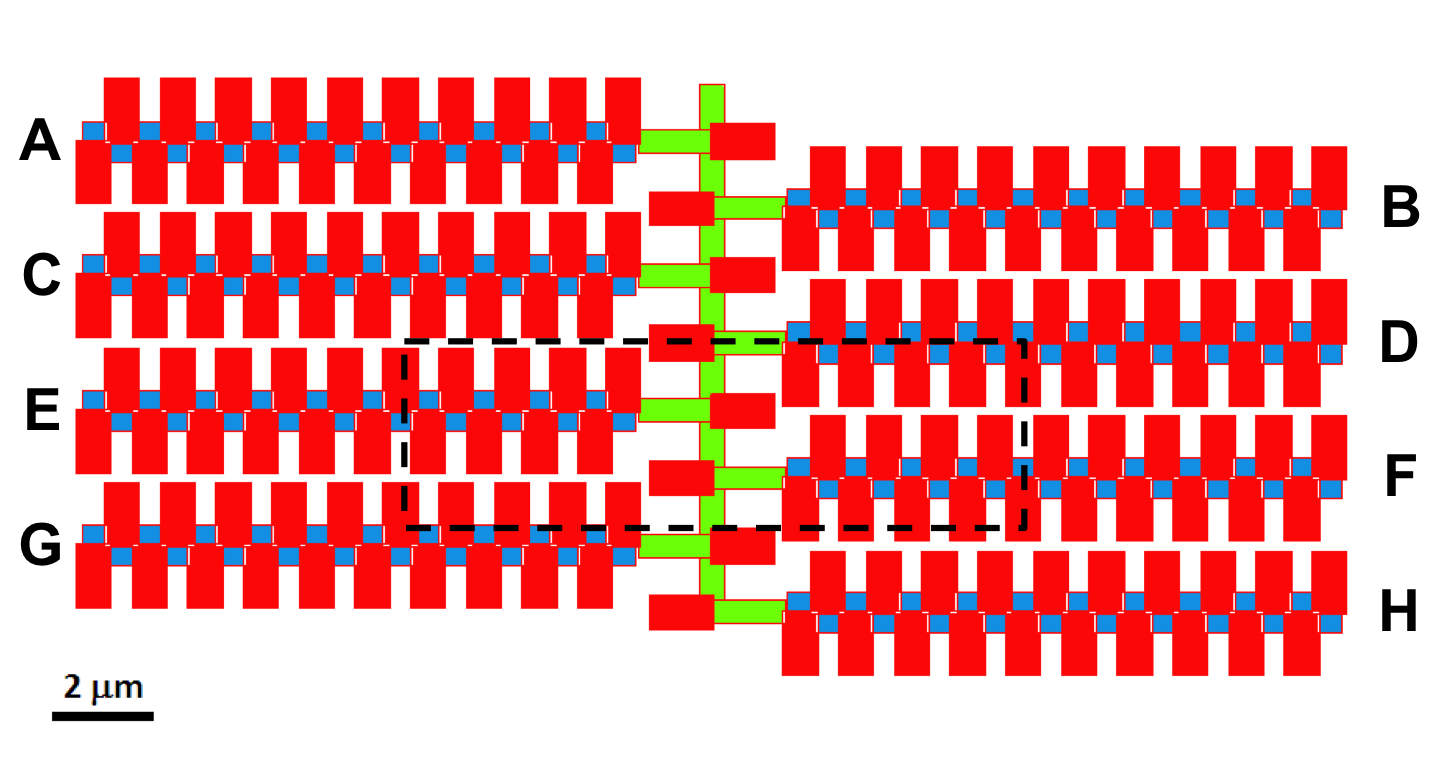}
\vspace*{13pt} \fcaption{Lithographic mask of a quantum register made of 8 logic qubits (A-H).
Every logical qubit is composed of 20 double data qubits (see the right side of Figure \ref{fig:qubit}) depicted as blue boxes and 20 T blocks (see Figure \ref{fig:chain}) colored in red.
Connections between logic qubits are provided by chain modules (see Figure \ref{fig:chain}) colored in green and 8 additional T modules.
Such logic quantum byte is composed of 1720 data qubits and 1400 communication qubits and covers an area of 307.502 $\mu$m$^2$}
\label{fig:Log_qubyte}       
\end{figure}

In the bus structure of Figure \ref{fig:Log_qubyte}, one of the 8 branches is a logic qubit according to the $[[7,1,3]]$ code and it is composed of 20 physical data qubits, including \emph{ancillae} qubits.
Tab. \ref{Tab:Area} reports the dimensions and compositions of the principal quantum gates and multi-qubit circuits presented in this work.
In particular, the 8-qubit block in Figure \ref{fig:Log_qubyte} is the quantum analog of the classical unit of information (the byte) and can be taken as a first benchmark to estimate the maximum density of logic qubits per unit area.
Such register covers an active area of 25.54 x 12.04 = 307.502 $\mu$m$^2$, that corresponds to a density of information of 2.6 Mqubit per cm$^2$.

\begin{figure}[h!]
\centering
  \includegraphics[width=0.85\columnwidth]{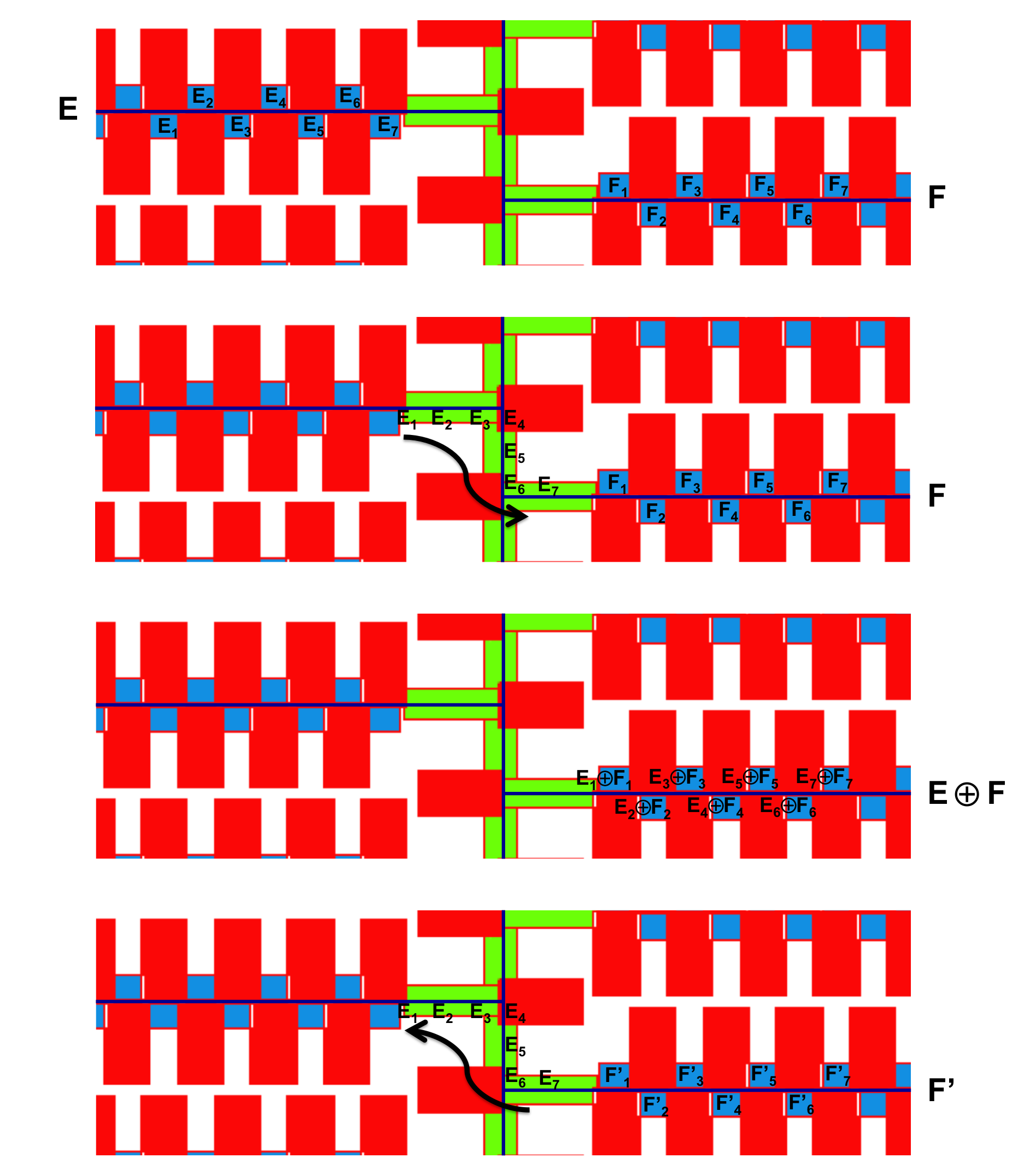}
\vspace*{13pt} \fcaption{Operation of a two-qubit logic gate between encoded qubits. The reported mask refers to the box highlighted in Figure \ref{fig:Log_qubyte}.
A logical qubit (E) is transferred in proximity of a second logical qubit (F) by sequencial SWAP operations through communication qubits.
Then a two-qubit logic gate ($e.g.$ CNOT) is operated between all the couples of physical qubits ($E_i\oplus F_i$) within data qubit blocks, effectively carrying out a CNOT gate between the logic qubits ($E\oplus F$).
The target qubit F is modified by the CNOT gate according to the state of the control qubit E, resulting in a new quantum state F'.
In the end the two qubits are moved to other qubit registers for further logic operations.
}
\label{fig:Operation}       
\end{figure}
A remarkable property of the $[[7,1,3]]$ code is that the fundamental logic gates operate on logic qubits in complete analogy to logic gates on physical qubits.
More in detail, the operation principle of a logic gate between two encoded qubits A and B is sketched in Figure \ref{fig:Operation}.
Logic qubit A is firstly transferred through a SWAP channel in proximity to qubit B, where the logic gate is carried out on a one-by-one basis between all the couples of physical qubits.
Finally qubit A is brought back to the starting point or directed to another qubit site for the next step in the algorythm.
Notably, all the logic operations are performed fault-tolerantly within this scheme, $i. e.$ interaction between physical qubits in the same logic qubit never takes place.
As a consequence, the propagation of errors inside a logic qubit is forbidden, preserving the possibility to perform quantum error correction over single faults.


\begin{table}[h!]
\tcaption{\label{Tab:Area}
Physical dimensions and composition of data and communication qubits with $d_{iD}$ = 40 nm. The last two rows report the dimensions of a logic qubit and of a register of 8 logic qubits respectively.
}
\centerline{\footnotesize\smalllineskip
\begin{tabular}{@{}ccccc}
\toprule
Device & Dimensions [$\mu$m$^2$] & Area [$\mu$m$^2$] & Data/Comm. qubits\\
\midrule
One-qubit & 0.3 x 0.5 & 0.15 & 1/0\\ 
Two-qubit & 0.38 x 0.5 & 0.19 & 2/0\\ 
Chain & 0.16 x 0.46 & 0.0736 & 0/2\\ 
T & 1.3 x 0.7 & 0.91 & 0/7\\ 
1 Log. qubit & 11.38 x 2.52 & 28.6776 & 20/70\\ 
8 Log. qubits & 25.54 x 12.04 & 307.502 & 1720/1400\\ 
\bottomrule
\end{tabular}}
\end{table}

In order to evaluate the physical performances of this architecture, we report in Tab. \ref{Tab:Times} the path length between two adjacent physical/logic qubits to estimate the corresponding time needed for quantum communication.
According to the analysis in Subsection \ref{subsec:comm_qubits}, $t_\text{SWAP}$ = 6.47 ns for $d_\text{iD}=40$ nm.
As a result, the time needed to transfer quantum information through a SWAP chain ranges from 71.2 ns for communication between adjacent physical qubits to approximately 2 $\mu$s for coherent transfer between logic qubits within a 8-qubit register.

\begin{table}[h!]
\tcaption{\label{Tab:Times}
Time load for quantum computation and communication between distant data qubits with a 40 nm inter-QD distance.
According to the analysis in Subsection \ref{subsec:comm_qubits} $t_{ \text{SWAP} }$ is 6.47 ns. 
The minimum and maximum transfer times for quantum communication have been calculated considering the minimum and maximum distance between physical qubits in the same logic qubit (made of 20 physical qubits) and between different logic qubits in a byte (8 logic bits).
}
\centerline{\footnotesize\smalllineskip
\begin{tabular}{@{}cccc}
\toprule
Operation & Number of Qubits & Distance [$\mu$m] & Time [ns] \\
\midrule
Comm. 2 phys. qubits (min) & 12 & 1 & 71.2 \\ 
Comm. 2 phys. qubits (max) & 138 & 11 & 886.4 \\ 
Comm. 2 log. qubits (min) & 192 & 15.4 & 1235.8 \\ 
Comm. 2 log. qubits (max) & 311 & 24.9 & 2005.7 \\ 
\bottomrule
\end{tabular}}
\end{table}

The estimated characteristic times for quantum information processing should be compared to the qubit coherence time ${T_2}^*$.
Although the first experimental works reported a short ${T_2}^* \sim$ 20 ns for a silicon hybrid qubit \cite{Shi_Nature_Hybrid,Hybrid_2014}, the expected value from theoretical calculations is of the order of $\mu$s \cite{Shi_Hybrid}.
A fidelity of 99.999\% seems to be within range for 1-qubit operations and could be further improved by tuning the singlet-triplet splitting to a good balance between operational speed and gate fidelity \cite{Shi_Hybrid,Koh_PNAS}.
Besides this, the effects of the principal sources of decoherence could be drastically reduced with several techniques, such as dynamical decoupling and complex pulse sequencies \cite{Bluhm_Dephasing}.
Finally, promising alternatives could be considered to replace the SWAP channels, expecially for quantum communication over long distances, such as teleportation gates and coherent transfer by adiabatic passage \cite{IEEE_SiP_Arch,CTAP_Greentree,CTAP_Platero,Bennett_Teleport}.

We also note that the bus-structure reported in Figure \ref{fig:Log_qubyte} can be easily extended to higher order ramifications in order to introduce recursive coding techniques \cite{QEC_Beginners}.
A recursive code of order $k-1$ gives an error threshold of $(cp)^{2^k} /c$ where $p$ is the error probability of a logic operation and 1/$c$ is the error threshold, $i. e.$ the maximum error rate tolerated by a specific quantum code \cite{QEC_Beginners,IEEE_SiP_Arch}.
As a result, recursive coding rises the error threshold by re-encoding logical qubits in a higher level logic qubit provided that $p < 1/c$. 
If this condition is satisfied, the error threshold is enhanced by an exponential law, whereas the circuit area and the computational times increase by a power law \cite{IEEE_SiP_Arch}.

\section{Conclusions}
\noindent
A CMOS-compatible design of the semiconductor hybrid qubit architecture has been proposed.
Such architecture is suitable for large scale quantum computing, since it allows all-electrical manipulation of qubits on a nanosecond timescale.

One- and two-qubit gates have been designed for a Si-CMOS platform, complying with the technologic standards of semiconductor industry.
The fundamental building blocks for quantum computation and communication have been proposed, and the feasability of multi-qubit networks has been discussed.
The requirements of fault-tolerant computation and the introduction of a quantum error correction scheme based on the $[[7,1,3]]$ Steane code have been taken into account.
The time and space resources for universal quantum computation are estimated accordingly in a register of 8 logical qubits.

The calculated maximum surface density of logical qubits is 2.6 Mqubit/cm$^2$.


\nonumsection{References}
\noindent

\bibliographystyle{spmpsci}      
\bibliography{Architecture_Hybrid_Qubit}   
%
%
%
%
%
%

\end{document}